\documentclass[aps,pra,twocolumn,showpacs]{revtex4}
\usepackage{graphicx}
\usepackage{psfrag}
\usepackage{amsmath}
\usepackage{amssymb}
\usepackage{bm}

\renewcommand{\vec}[1]{\boldsymbol{#1}}
\renewcommand{\tensor}[1]{{\mathsf {#1}}}
\newcommand{\Nabla}{\vec{\nabla}}
\newcommand{\Lnabla}{\overset{\leftarrow}{\Nabla}}

\newcommand{\EW}[1]{\langle #1 \rangle}

\newcommand{\D}{\mathrm{d}}
\newcommand{\bdelta}{\Delta}
\renewcommand{\Im}{{\rm Im}\,}

\newcommand{\uv}[1]{\mathbf{e}_{#1}}

\newcommand{\fo}[1]{\underline{\hat{\mathbf{#1}}}}

\begin{document}
\title{Dispersive forces on bodies and atoms:
a unified approach}
\author{Christian Raabe}
\author{Dirk-Gunnar Welsch}
\affiliation{Theoretisch-Physikalisches Institut,
Friedrich-Schiller-Universit\"at Jena, Max-Wien-Platz 1, D-07743 Jena,
Germany}
\date{\today}
%%%%%%%%%%%%%%%%%%%%%%%%%%%%%%%%%%%%%%%%%%%%%%%%%%%%%%%%%%%%%%%%%%%%%%%
\begin{abstract}
A unified approach to the calculation of dispersive forces
on ground-state bodies
and atoms is given. It is based on
the ground-state Lorentz force density acting on the charge
and current densities
attributed to the polarization and magnetization in linearly,
locally, and causally responding media. The theory is applied to
dielectric macro- and micro-objects, including single atoms.
Existing formulas valid for weakly polarizable matter
are generalized to allow also for strongly polarizable matter.
In particular when micro-objects can be regarded
as single atoms, well-known formulas for the Casimir-Polder force
on atoms and the van der Waals interaction between atoms
are recovered. It is shown that the force acting on 
medium atoms---in contrast to isolated atoms---is in general 
screened by the other medium atoms.
\end{abstract}
\maketitle
%%%%%%%%%%%%%%%%%%%%%%%%%%%%%%%%%%%%%%%%%%%%%%%%%%%%%%%%%%%%%%%%%%%%%%%
\section{Introduction}
\label{sec1}
It is well known that polarizable particles and
macroscopic bodies---matter whose electromagnetic properties
are described terms of macroscopic quantities---are subject 
to forces in the presence of
electromagnetic fields, even if the (fluctuating) fields vanish
on average and the bodies do not carry any excess charges and 
are unpolarized. In particular, this is also the case when
the field-matter system can be assumed to be in its ground state,
where only quantum fluctuations are responsible for the forces.
In this case it is common to speak of vacuum forces or dispersive 
forces, which obviously represent a genuine quantum effect.
Basically, it can be distinguished between three kinds
of dispersive forces, namely van der Waals (vdW), Casimir-Polder (CP), 
and Casimir forces, depending on whether forces between individual 
particles, between particles and macroscopic bodies or between 
macroscopic bodies are respectively considered
(see, e.g., Ref.~\cite{MilonniQuantumVacuum} for an overview).

Since both CP forces on atoms and Casimir forces on macroscopic
bodies may be regarded as being macroscopic manifestations
of microscopic van der Waals forces, intimate relations between
them can be expected. Nevertheless, quite different theoretical
approaches to the two kinds of forces have been developed.
Compared to the large body of work in this field, only little
attention has been paid to their common origin and consequential
relations between them (see, e.g.,
Refs.~\cite{SchwingerJ091978,TomasMS022005,TomasMS092005,
TomasMS002005,RaabeC112005}).
Moreover, the studies have been based on specific geometries
such as simple planar structures, and weakly polarizable matter 
has been considered.

More attention has been paid to the relations between Casimir 
forces and van der Waals forces, but again for
specific geometries and weakly polarizable matter (see, e.g., 
Refs.~\cite{CasimirHBG021948,LifshitzEM001955,AbrikosovStatPhys,
SchwingerJ091978,MiltonKA051998,
BrevikI051999,BartonG001999}). Quite recently, a very general 
relation between CP forces (as calculated within the
frame of macroscopic QED) and multi-atom van der Waals forces 
has been established \cite{BuhmannSY102005}, where it
has been shown that the CP force acting on an atom in the presence
of a dielectric body of given permittivity can be regarded as being
the sum of all many-atom van der Waals forces
with respect to the atoms of the body.

In this paper we develop, within the framework of QED in
linearly, locally, and causally responding magnetodielectric media, 
a unified approach to the calculation of
dispersive forces acting on ground-state macro- and 
micro-objects. Since the origin of any electromagnetic force
is the Lorentz force acting on appropriate charges and currents,
we first consider the ground-state expectation value of the
Lorentz force density acting on the charge and current
densities attributed to the polarization and magnetization fields
of linear media, taking fully into account the noise polarization
and noise magnetization that are associated with absorption.
From the ground-state Lorentz force density obtained in this way,
the force acting on an arbitrary body or an arbitrary part of it
can be then obtained by integration over the respective
volume. Applying the theory to dielectric systems, we present a very
general force formula, the applicability of which ranges from
dielectric macro-objects to micro-objects, also including single 
atoms, without restriction to weakly dielectric material.
In particular, this formula enables us to extend the well-known 
CP-type formula for the force acting on a weakly dielectric 
(mic\-r\mbox{o-)}object to an arbitrary one.

The paper is organized as follows.
In Sec.~\ref{sec2} the theory of CP forces acting on single
ground-state atoms in the presence of magnetodielectric
bodies is 
recapitulated. The ground-state expectation value of the Lorentz 
force density in linear magnetodielectric media is calculated 
in Sec.~\ref{sec3}. On this basis, in Sec.~\ref{sec4} the force 
acting on an arbitrary dielectric body or a part of it is derived.
The theory is applied in Sec.~\ref{sec5} to the calculation
of the force acting on a dielectric micro-object, the limiting
case of a single atom (which can be either an isolated atom or a 
medium atom) is considered, and contact to earlier results found 
for planar structures is made. In Sec.~\ref{sec6} it is shown that
the theory can also be used to study the van der Waals interaction
between two atoms. Finally, a summary and some concluding remarks 
are given in Sec.~\ref{sec7}. For the sake of clarity, some of the
derivations are given in appendices.
%
%%%%%%%%%%%%%%%%%%%%%%%%%%%%%%%%%%%%%%%%%%%%%%%%%%%%%%%%%%%%%%%
%%%%%%%%%%%%%%%%%%%%%%%%%%%%%%%%%%%%%%%%%%%%%%%%%%%%%%%%%%%%%%%
%
\section{Casimir-Polder force}
\label{sec2}
The CP force acting on
a ground-state atom in the vicinity of arbitrary
magnetodielectric bodies that linearly, locally, and causally
respond to the electromagnetic field can be
regarded as being a conservative force. Hence it can be
given by the negative gradient of a potential which in
the leading order of perturbation theory reads 
(see, e.g., \cite{McLachlanAD001963,AgarwalGS001975,
WylieJM001984,HenkelC002002,BuhmannSY032004})
\begin{equation}
\label{L6}
\mathbf{F}^{\mathrm{(at)}}
(\mathbf{r})=
-
\frac{\hbar\mu_{0}}{2\pi}
\int_{0}^{\infty}\D\xi\,
\xi^2 \alpha(i\xi)
\Nabla
\mathrm{Tr\,}\tensor{G}^{\mathrm{(S)}}
(\mathbf{r},\mathbf{r},i\xi),
\end{equation}
where $\mathbf{r}$ is the position of the atom,
$\alpha(i\xi)$ is its polarizability, and
$\tensor{G}^{\mathrm{(S)}}(\mathbf{r,r'},i\xi)$
is the scattering part of the classical retarded Green tensor
$\tensor{G}(\mathbf{r,r'},i\xi)$ taken at imaginary frequencies.
Note that the scattering part of the Green tensor contains all the
necessary information about the configuration of the
magnetodielectric bodies, whose electromagnetic properties
are characterized by the electric and magnetic susceptibilities 
that are complex functions of frequency and may vary in space.
The (translationally invariant) bulk part of the Green tensor,
which would diverge in the coincidence limit, is not needed 
in Eq.~(\ref{L6}), because it cannot contribute to the force.
Equation (\ref{L6}), which can be derived \cite{BuhmannSY032004}
on the basis of exact quantization of the macroscopic
electromagnetic field in linearly, locally, and causally 
responding media \cite{KnoellL002001}, strictly applies to isolated
atoms, but not to medium atoms nor to guest atoms in a substrate 
medium. Note that the atomic ground-state
polarizability $\alpha(\omega)$ in leading-order
perturbation theory,
\begin{equation}
\label{L5a}
\alpha(\omega)
\sim\sum_{k} \frac{\Omega_{k}^2}{\omega_{k}^2-\omega^2}\,
,
\end{equation}
features poles on the real frequency axis due to the neglect of 
level broadening. If necessary, the correct response function 
properties \cite{LanLif} may be restored by means of an appropriate 
limit prescription, viz.
\begin{equation}
\label{L5b}
\alpha(\omega)
\sim
\lim_{\gamma\to 0+}
\sum_{k} \frac{\Omega_{k}^2}{\omega_{k}^2-\omega^2-i\gamma\omega}\,.
\end{equation}

In order to apply Eq.~(\ref{L6}) to a dielectric body
of volume $V_\mathrm{M}$, let us consider instead of a single atom
a collection of atoms that are (strictly) contained inside
a space region of volume
$V_\mathrm{M}$, and let us add up the individual forces as given
by Eq.~(\ref{L6}). Since the mutual interaction of the atoms
is completely ignored in this way, it is clear that this method 
gives only the lowest-order approximation to the total force.
If the number density of the atoms (defined on a suitably chosen 
macroscopic length scale) is denoted by $\eta(\mathbf{r})$, the 
total force in this approximation reads
\begin{equation}
\label{L8}
\mathbf{F}
= -\frac{\hbar\mu_{0}}{2\pi}
\int_{V_\mathrm{M}} \D^3r
\int_{0}^{\infty}\D\xi\,
\xi^2
\eta
(\mathbf{r})\alpha
(i\xi)
\Nabla
\mathrm{Tr\,}\tensor{G}^\mathrm{(S)}
(\mathbf{r},\mathbf{r},i\xi).
\end{equation}
Since the validity of Eq.~(\ref{L8}) obviously requires
sufficiently weakly polarizable atoms and/or a sufficiently low
number density of atoms, the collection of atoms can be viewed as
dielectric matter of volume $V_\mathrm{M}$
and small susceptibility
\begin{equation}
\label{L9}
\chi_\mathrm{M}(\mathbf{r},i\xi)
=\eta(\mathbf{r})\alpha(i\xi)
/\varepsilon_{0},
\end{equation}
which implies that the permittivity of the overall system
has slightly been changed by $\delta\varepsilon(\mathbf{r},i\xi)$
$\!=$ $\!\chi_\mathrm{M}(\mathbf{r},i\xi)$. 
In particular, applying Eq.~(\ref{L8}) to a dielectric micro-object
whose number density of atoms is constant over the small
volume $V_\mathrm{M}$, we obtain the force
\begin{equation}
\label{L8b}
\mathbf{F}
= - V_\mathrm{M}\eta\,\frac{\hbar\mu_{0}}{2\pi}
\int_{0}^{\infty}\D\xi\,
\xi^2
\alpha(i\xi)
\Nabla
\mathrm{Tr\,}\tensor{G}^\mathrm{(S)}
(\mathbf{r},\mathbf{r},i\xi).
\end{equation}
Clearly, application of Eq.~(\ref{L8}) to dielectric bodies
(including micro-objects) that are dense and/or consist
of strongly polarizable atoms becomes questionable.
Hence, the problem of a generalization of Eq.~(\ref{L8})
to arbitrary dielectric bodies or parts of them arises.
As we will see, a satisfactory answer can be given
on the basis of the ground-state Lorentz force
density in media.

%%%%%%%%%%%%%%%%%%%%%%%%%%%%%%%%%%%%%%%%%%%%%%%%%%%%%%%%%%%%%%%%%
%
\section{Ground-state Lorentz force}
\label{sec3}
As shown in Ref.~\cite{RaabeC012005}, the Casimir force between
(linearly, locally and causally responding) macroscopic bodies 
can be regarded as the expectation value of the Lorentz force 
acting on the charges and currents attributed to the polarization 
and magnetization of the bodies. To be more specific, the
charge and current densities $\hat{\rho}(\mathbf{r})$ and
$\hat{\mathbf{j}}(\mathbf{r})$, respectively, which the Lorentz 
force density acts on are given by
\begin{equation}
\label{L40}
\hat{\rho}(\mathbf{r})=\int_{0}^\infty
\D\omega\,
\fo{\rho}(\mathbf{r},\omega) + \mathrm{H.\,c.}
\end{equation}
and $\hat{\mathbf{j}}(\mathbf{r})$ accordingly, with
\begin{equation}
\label{L40a}
\fo{\rho}(\mathbf{r},\omega) =
-\varepsilon_{0}
\Nabla\cdot\{[\varepsilon(\mathbf{r},\omega)-1]\fo{E}(\mathbf{r},\omega)\}
+ (i\omega)^{-1} \Nabla\cdot\fo{j}_\mathrm{N}(\mathbf{r},\omega)
\end{equation}
and [$\kappa_{0}\!=\!\mu_{0}^{-1}$]
\begin{align}
\label{L40b}
&\fo{j}(\mathbf{r},\omega)
=-i\omega \varepsilon_{0}
[\varepsilon(\mathbf{r},\omega)-1]
\fo{E}(\mathbf{r},\omega)
\nonumber\\&\quad
+ \Nabla\times \{\kappa_{0}[1-\kappa(\mathbf{r},\omega)]
\fo{B}(\mathbf{r},\omega)\}
+\fo{j}_\mathrm{N}(\mathbf{r},\omega).
\end{align}
Here, $\fo{j}_{\mathrm{N}}(\mathbf{r},\omega)$ is the (fluctuating) 
noise current density that acts as a Langevin noise source in the
(macroscopic) Maxwell equations in the $\omega-$domain, 
$\varepsilon(\mathbf{r},\omega)$ 
and $\kappa^{-1}(\mathbf{r},\omega)$ are respectively
the (complex) permittivity and permeability, and
$\fo{E}(\mathbf{r},\omega)$ and $\fo{B}(\mathbf{r},\omega)$ 
are the (positive) frequency components of the medium-assisted
electromagnetic field:
\begin{align}
\label{L43}
&\fo{E}(\mathbf{r},\omega)
=i\mu_{0}\omega\int \D^3r'\,
\tensor{G}(\mathbf{r,r'},\omega)
\cdot
\fo{j}_{\mathrm{N}}(\mathbf{r'},\omega),
\\[1ex]&
\label{L44}
\fo{B}(\mathbf{r},\omega)
=\mu_{0}\Nabla\times \int \D^3r'\,
\tensor{G}(\mathbf{r,r'},\omega)
\cdot
\fo{j}_{\mathrm{N}}(\mathbf{r'},\omega)
%.
\end{align}
(for details of the quantization scheme, see 
Ref.~\cite{KnoellL002001}). The expectation value of the Lorentz
force,
\begin{multline}
\label{L45}
\mathbf{F}=\int_{V_\mathrm{M}} \D^3r\,
\int_{0}^\infty \D\omega\,\int_{0}^\infty \D\omega'\,
\EW{\fo{\rho}(\mathbf{r},\omega)
\fo{E}{^{\dagger}}(\mathbf{r'},\omega')
\\
+\fo{j}(\mathbf{r},\omega)
\times\fo{B}{^{\dagger}}(\mathbf{r'},\omega')}_{\mathbf{r'}
\to\mathbf{r}},
\end{multline}
taken with respect to the ground state of the linearly interacting 
field-matter system then yields the Casimir force acting
in the zero-temperature limit on the material in a chosen spatial 
region of volume $V_\mathrm{M}$. Again, divergent bulk contributions 
must be discarded in the limit $\mathbf{r'}\!\to\!\mathbf{r}$,
as they would correspond to an unphysical ``self-forces" of the
respective volume elements.

The Casimir force as given by Eq.~(\ref{L45}) [together with
Eqs.~(\ref{L40a})--(\ref{L44})] can be equivalently rewritten as 
a surface integral over a stress tensor,
\begin{equation}
\label{L45-1}
\mathbf{F}=\int_{\partial V_\mathrm{M}}
\D\mathbf{a}
\cdot\tensor{T}(\mathbf{r}),
\end{equation}
where (at zero temperature)
\begin{equation}
\label{L12}
\tensor{T}(\mathbf{r})=
\lim_{\mathbf{r}'\to\mathbf{r}}\left[
\tensor{S}(\mathbf{r,r'})
- {\textstyle\frac{1}{2}} \tensor{1}
{\rm Tr\,}\tensor{S}(\mathbf{r,r'})
\right]
\end{equation}
together with
\begin{align}
\label{L13}
\tensor{S}(\mathbf{r,r'})
&
=-\frac{\hbar}{\pi}\int_{0}^{\infty} \D\xi\,
\left[\frac{\xi^2}{c^2}\,
\tensor{G}^{\mathrm{(S)}}(\mathbf{r,r'},i\xi)
\right.
\nonumber\\&\qquad\quad
\left.
+\Nabla\times
\tensor{G}^{\mathrm{(S)}}(\mathbf{r,r'},i\xi)
\times\Lnabla{'}
\right]
.
\end{align}
Here and in the following, $\tensor{1}$ denotes the unit 
tensor and $\times\Lnabla{'}$ is meant to act to its left.
With regard to the controversial view held in 
Ref.~\cite{PitaevskiiL??2006}, it should be emphasized that
Eq.~(\ref{L45-1}) together with Eqs.~(\ref{L12}) and (\ref{L13}) 
yields the genuine electromagnetic Casimir force acting on bodies
or pieces of them. Clearly, in mechanical equilibrium
the Casimir force is balanced by additional internal
or external (mechanical) forces that are not included in
the equations considered here.

Let us return to Eq.~(\ref{L45}) together with 
Eqs.~(\ref{L40a})--(\ref{L44}). It is not difficult to show 
(Appendix \ref{AppA}) that Eqs.~(\ref{L40a}) and (\ref{L40b})
can be rewritten as
\begin{align}
\label{L41}
&\fo{\rho}(\mathbf{r},\omega)
=\frac{i\omega}{c^2}\, \Nabla \cdot\int \D^3r'\,
\tensor{G}(\mathbf{r,r'},\omega)
\cdot
\fo{j}_{\mathrm{N}}(\mathbf{r'},\omega),
\\[1ex]&
\label{L42}
\fo{j}(\mathbf{r},\omega)
=\left(\Nabla\times\Nabla\times
-\frac{\omega^2}{c^2}\right)
\int \D^3r'\,
\tensor{G}(\mathbf{r,r'},\omega)
\cdot
\fo{j}_{\mathrm{N}}(\mathbf{r'},\omega),
\end{align}
and that the relation
\begin{align}
\label{L51}
&\EW{\fo{j}_{\mathrm{N}}(\mathbf{r},\omega)
\fo{j}_{\mathrm{N}}^{\dagger}(\mathbf{r'},\omega')}
\nonumber\\&\qquad
=\frac{\hbar}{\mu_{0}\pi}\,\delta(\omega-\omega')
\biggl\{
\frac{\omega^2}{c^2}\,
\Im\,\varepsilon(\mathbf{r},\omega)
\tensor{1}\delta(\mathbf{r-r'})
\nonumber\\&\qquad\quad
- \Nabla\times[\Im\,\kappa(\mathbf{r},\omega)
\tensor{1}\delta(\mathbf{r-r'})]\times\Lnabla{'}
\biggr\}
\end{align}
holds for the ground-state expectation value. Employing standard 
properties of the Green tensor (see, e.g., Ref.~\cite{KnoellL002001}),
from Eqs.~(\ref{L43}), (\ref{L44}) and (\ref{L41})--(\ref{L51})
it follows that
\begin{equation}
\label{L51a}
\EW{\fo{\rho}(\mathbf{r},\omega)\fo{E}{^{\dagger}}(\mathbf{r'},\omega')}=
\frac{\hbar}{\pi}\frac{\omega^2}{c^2}\,\delta(\omega-\omega')
\Nabla\cdot\Im \tensor{G}(\mathbf{r,r'},\omega)
\end{equation}
and
\begin{align}
\label{L51b}
&\EW{\fo{j}(\mathbf{r},\omega)\fo{B}{^{\dagger}}(\mathbf{r'},\omega')}
\nonumber\\&\
=-\frac{\hbar}{\pi}\,\delta(\omega-\omega')
\biggl(\Nabla\times\Nabla\times
-\frac{\omega^2}{c^2}\biggr)
\Im \tensor{G}(\mathbf{r,r'},\omega)\times\Lnabla{'},
\end{align}
which can be used to express the integrand of the volume
integral in Eq.~(\ref{L45})---that is, the Casimir
force density---in terms of the (scattering part of the)
Green tensor solely. Extension of the theory to include
thermal states is straightforward.

%%%%%%%%%%%%%%%%%%%%%%%%%%%%%%%%%%%%%%%%%%%%%%%%%%%%%%%%%%%%%%%%%
\section{Force on dielectric bodies}
\label{sec4}
\begin{figure}[htb]
\psfrag{A}[][]{$V$}
\psfrag{B}[][]{M}
\psfrag{C}[][]{$V_\mathrm{M}$}
\psfrag{D}[][]{$V_\mathrm{B}$}
\includegraphics[width=0.9\linewidth]{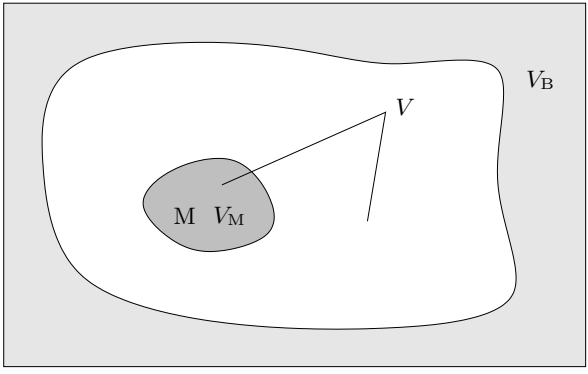}
\caption{
\label{Fig1}
A dielectric body M of volume $V_\mathrm{M}$
inside an empty-space region of total volume $V$. 
There may be arbitrary magnetodielectric bodies in
the outer region $V_\mathrm{B}$.
}
\end{figure}

\begin{figure}[htb]
\psfrag{A}[][]{$V$}
\psfrag{B}[][]{M}
\psfrag{C}[][]{$V_\mathrm{M}$}
\psfrag{D}[][]{$V_\mathrm{B}$}
\includegraphics[width=0.9\linewidth]{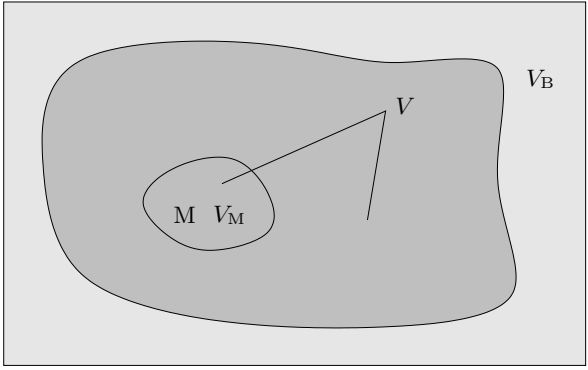}
\caption{
\label{Fig2}
A dielectric body M of volume $V_\mathrm{M}$ that is an inner 
part of a larger dielectric body of volume $V$. There may
be arbitrary magnetodielectric bodies in the outer region 
$V_\mathrm{B}$.
}
\end{figure}

To illustrate the theory, we first apply Eq.~(\ref{L45}) 
[together with Eqs.~(\ref{L51a}) and (\ref{L51b})] to the 
calculation of the Casimir force acting on a dielectric body 
in some space region of volume $V$. We want to 
study---mostly in parallel---the two cases sketched in 
Figs.~\ref{Fig1} and \ref{Fig2}, namely (i) an isolated body 
(Fig.~\ref{Fig1}), and (ii) a body that is an inner part of 
some larger body (Fig.~\ref{Fig2}). In both cases, arbitrary
magnetodielectric bodies are allowed to be present in the 
outer region $V_\mathrm{B}$ in the figures.
%
%%%%%%%%%%%%%%%%%%%%%%%%%%%%%%%%%%%%%%%%%%
%
%%%%%%%%%%%%%%%%%%%%%%%%%%%%%%%%%%%%%%%%%%%%%%%%%%%%

Let $\varepsilon(\mathbf{r},\omega)$ be the permittivity
of the system in the absence of the dielectric matter
in $V$ and assume that it changes to $\varepsilon(\mathbf{r},\omega)$ 
$\!+$ $\!\bdelta\varepsilon(\mathbf{r},\omega)$ when the 
additional dielectric matter is introduced into the initially empty
space region $V$. Here and in the following, $\bdelta{A}$
denotes the exact change of a quantity $A[\varepsilon(\mathbf{r},\omega)]$
due to a given (not necessarily small) 
change of the permittivity, $\bdelta{\varepsilon}
(\mathbf{r,\omega})$, whereas the notation $\delta{A}$
is used to indicate the familiar first-order variation 
of $A$ produced by a small variation
$\delta{\varepsilon}(\mathbf{r},\omega)$.
Further, let $\tensor{G}_{V}(\mathbf{r,r'},\omega)$ and
$\tensor{G}(\mathbf{r,r'},\omega)$ be the Green tensors
of the system in the cases where the dielectric matter inside 
the space region $V$ is present and absent, respectively,
with both of them taking into account the magnetodielectric bodies
in the space region $V_\mathrm{B}$ in Figs.~\ref{Fig1} and \ref{Fig2}.
We may then write
\begin{equation}
\label{L14c}
\tensor{G}_{V}
(\mathbf{r,r'},\omega)=\tensor{G}
(\mathbf{r,r'},\omega)+\bdelta\tensor{G}
(\mathbf{r,r'},\omega),
\end{equation}
where the change of the Green tensor,
$\bdelta\tensor{G}(\mathbf{r,r'},\omega)$, due to
the introduction of the dielectric matter into $V$
obeys the Dyson-type equation
\begin{equation}
\label{L14a}
\bdelta\tensor{G}
(\mathbf{r,r'},\omega)=\frac{\omega^2}{c^2}\int \D^3s\,
\tensor{G}(\mathbf{r,s},\omega)\cdot
\bdelta\varepsilon(\mathbf{s},\omega)
\tensor{G}_{V}(\mathbf{s,r'},\omega),
\end{equation}
with the integral running over the space region $V$ (at most).

To calculate the changes of the expectation values
$\EW{\fo{\rho}(\mathbf{r},\omega)\fo{E}{^{\dagger}}(\mathbf{r'},\omega')}$
[Eq.~(\ref{L51a})] and
$\EW{\fo{j}(\mathbf{r},\omega)\fo{B}{^{\dagger}}(\mathbf{r'},\omega')}$
[Eq.~(\ref{L51b})] ($\mathbf{r}$ $\!\in$ $\!V$), we note that for
$\mathbf{r}$ $\!\in$ $\!V$ the Green tensor $\tensor{G}(\mathbf{r,r'},
\omega)$ satisfies the same differential equation as the
free-space Green tensor, i.e.,
\begin{equation}
\label{L57}
\left(\Nabla\times\Nabla\times
-\frac{\omega^2}{c^2}\right)
\tensor{G}(\mathbf{r,r'},\omega)=
\tensor{1}\delta(\mathbf{r-r'})
\qquad (\mathbf{r}\in V),
\end{equation}
from which it follows that
\begin{equation}
\label{L53}
\frac{\omega^2}{c^2}
\Nabla\cdot\tensor{G}(\mathbf{r,r'},\omega)
=-\Nabla\delta(\mathbf{r-r'})
\qquad (\mathbf{r}\in V).
\end{equation}
Hence from Eqs.~(\ref{L14a}) and (\ref{L53}) we derive
\begin{align}
\label{L53a}
&\Nabla\cdot\Im\bdelta\tensor{G}
(\mathbf{r,r'},\omega)
\nonumber\\&\quad
=
-\Nabla\cdot\Im\left[
\bdelta\varepsilon(\mathbf{r},\omega)\tensor{G}_{V}
(\mathbf{r,r'},\omega)\right]
\qquad (\mathbf{r}\in V),
\end{align}
and thus from Eq.~(\ref{L51a}) the change
$\bdelta\EW{\fo{\rho}(\mathbf{r},\omega)\fo{E}{^{\dagger}}
(\mathbf{r'},\omega')}$ is found to be
\begin{align}
\label{L53b}
&\bdelta\EW{\fo{\rho}(\mathbf{r},\omega)
\fo{E}{^{\dagger}}(\mathbf{r'},\omega')}
=-
\frac{\hbar}{\pi}\,\frac{\omega^2}{c^2}\,
\delta(\omega-\omega')
\nonumber\\&\quad
%not a cross product!
\times\,
\Nabla\cdot\Im\left[
\bdelta\varepsilon(\mathbf{r},\omega)\tensor{G}_{V}
(\mathbf{r,r'},\omega)\right]
\qquad (\mathbf{r}\in V).
\end{align}
The change $\bdelta\EW{\fo{j}(\mathbf{r},\omega)\fo{B}{^{\dagger}}
(\mathbf{r'},\omega')}$ ($\mathbf{r}$ $\!\in$ $\!V$) can be found 
in a similar way. Making use of Eqs.~(\ref{L14a}) and (\ref{L57}),
from Eq.~(\ref{L51b}) we obtain
\begin{align}
\label{L51c}
&\bdelta\EW{\fo{j}(\mathbf{r},\omega)\fo{B}{^{\dagger}}(\mathbf{r'},\omega')}
=-\frac{\hbar}{\pi}\frac{\omega^2}{c^2}\,\delta(\omega-\omega')
\nonumber\\&\quad
%not a cross product!
\times\,
\Im[\bdelta\varepsilon(\mathbf{r},\omega)
\tensor{G}_{V}(\mathbf{r,r'},\omega)]
%a cross product
\times\Lnabla{'}
\qquad (\mathbf{r}\in V),
\end{align}
which implies that
\begin{align}
\label{L51d}
&\bdelta\EW{\fo{j}(\mathbf{r},\omega)\times\fo{B}{^{\dagger}}
(\mathbf{r'},\omega')}
\nonumber\\&\quad
=\frac{\hbar}{\pi}\frac{\omega^2}{c^2}\,\delta(\omega-\omega')
\Im\{\bdelta\varepsilon(\mathbf{r},\omega)\Nabla{'}
\mathrm{Tr\,}[\tensor{G}_{V}(\mathbf{r,r'},\omega)]
\nonumber\\&\qquad
-\bdelta\varepsilon(\mathbf{r},\omega)\Nabla{'}\cdot
\tensor{G}_{V}(\mathbf{r,r'},\omega)\}
\qquad (\mathbf{r}\in V).
\end{align}

Using Eqs.~(\ref{L53b}) and (\ref{L51d}), we can now easily
calculate, according to Eq.~(\ref{L45}), the Casimir force
acting on a dielectric body of volume $V_\mathrm{M}$ and 
dielectric susceptibility $\chi_\mathrm{M}(\mathbf{r},\omega)$
$\!\equiv$ $\!\bdelta\varepsilon(\mathbf{r},\omega)$
(cf. Figs.~\ref{Fig1} and \ref{Fig2}) as
\begin{align}
\label{L58}
&\mathbf{F}
=\frac{\hbar}{\pi c^2}
\int_{0}^{\infty} \D\omega\, \omega^2
\nonumber\\&\quad
\quad
\times\, \Im
\int_{V
_\mathrm{M}
} \D^3r\,
\bigl\{
\chi
_\mathrm{M}
(\mathbf{r},\omega)\Nabla{'}
\mathrm{Tr\,}[\tensor{G}_{V}(\mathbf{r,r'},\omega)]
\nonumber\\&\qquad
\quad
-(\Nabla+\Nabla{'})\cdot
\chi
_\mathrm{M}
(\mathbf{r},\omega)
\tensor{G}_{V}(\mathbf{r,r'},\omega)
\bigr\}_{\mathbf{r'}\to\mathbf{r}}.
\end{align}
The reciprocity property of
$\tensor{G}_{V}(\mathbf{r,r'},\omega)$
implies that
$\mathrm{Tr\,}\tensor{G}_{V}(\mathbf{r,r'},\omega)$
is symmetric with respect to $\mathbf{r}$ and $\mathbf{r'}$. 
Thus we may rewrite Eq.~(\ref{L58}) as
\begin{align}
\label{L59}
&
\mathbf{F}
=\frac{\hbar}{2\pi c^2}
\int_{0}^{\infty} \D\omega\,
\omega^2
\nonumber\\&\quad \times\,\biggl\{
\Im\! \int_{V
_\mathrm{M}
} \D^3r\,
\chi
_\mathrm{M}
(\mathbf{r},\omega)\Nabla
\mathrm{Tr\,}[
\tensor{G}_{V}(\mathbf{r,r'},\omega)
]
_{\mathbf{r'}\to\mathbf{r}}
\nonumber\\&\qquad\quad
- 2\, \Im\!\int_{\partial V
_\mathrm{M}
}
\D\mathbf{a}
\cdot
\chi
_\mathrm{M}
(\mathbf{r},\omega)
[\tensor{G}_{V}(\mathbf{r,r'},\omega)]
_{\mathbf{r'}\to\mathbf{r}}
\biggr\}.
\end{align}
Further, on recalling the analytic properties of the
integrands as functions of (complex) $\omega$, we may employ
contour integral techniques to represent Eq.~(\ref{L59})
in the form of
\begin{align}
\label{L60-1}
&
\mathbf{F}
=-\frac{\hbar}{2\pi c^2}
\int_{0}^{\infty} \D\xi\,
\xi^2
\nonumber\\&\quad \times\,\biggl\{
\int_{V
_\mathrm{M}
} \D^3r\,
\chi
_\mathrm{M}
(\mathbf{r},i\xi)\Nabla
\mathrm{Tr\,}[
\tensor{G}_{V}(\mathbf{r,r'},i\xi)]
_{\mathbf{r'}\to\mathbf{r}}
\nonumber\\&\qquad\quad
-2 \int_{\partial V
_\mathrm{M}
}
\D\mathbf{a}
\cdot
\chi
_\mathrm{M}
(\mathbf{r},i\xi)
[\tensor{G}_{V}(\mathbf{r,r'},i\xi)]
_{\mathbf{r'}\to\mathbf{r}}
\biggr\}.
\end{align}
In Eqs.~(\ref{L58})--(\ref{L60-1}) the coincidence limit
$\mathbf{r'}$ $\!\to$ $\!\mathbf{r}$ has again to be performed
in such a way that unphysical ``self-force'' contributions are
removed.

If the body consists of homogeneous dielectric matter, one can 
simply replace the Green tensor
$[\tensor{G}_{V}(\mathbf{r,r'},\omega)]_{\mathbf{r}'\to\mathbf{r}}$
with its scattering part
$\tensor{G}_{V}^\mathrm{(S)}(\mathbf{r},\mathbf{r},\omega)$.
In the case of inhomogeneous matter this replacement should be 
done point-wise. That is to say, at each space point
$\mathbf{r}$, the Green tensor for the corresponding bulk material 
must be subtracted from 
$\tensor{G}_{V}(\mathbf{r},\mathbf{r}',\omega)$
in the limit $\mathbf{r}'$ $\!\to$ $\!\mathbf{r}$.
In the case of an isolated body (cf. Fig.~\ref{Fig1}) the surface
integral in Eqs.~(\ref{L59}) and (\ref{L60-1}) can be dropped, hence
\begin{align}
\label{L60-2}
&
\mathbf{F}
=-\frac{\hbar}{2\pi c^2}
\int_{0}^{\infty} \D\xi\,
\xi^2
\nonumber\\&\quad \times\,
\int_{V
_\mathrm{M}
} \D^3r\,
\chi
_\mathrm{M}
(\mathbf{r},i\xi)\Nabla
\mathrm{Tr\,}[
\tensor{G}_{V}(\mathbf{r,r'},i\xi)]
_{\mathbf{r'}\to\mathbf{r}}.
\end{align}
Clearly, the surface integral must not be dropped
in the case where the
body is an inner part of a larger dielectric body
(cf. Fig.~\ref{Fig2}).

Equation (\ref{L60-2}) is the desired generalization of Eq.~(\ref{L8}).
In contrast to Eq.~(\ref{L8}), it represents the exact force acting on
a dielectric body of given permittivity, since
$\chi_\mathrm{M}(\mathbf{r},i\xi)$ is not
restricted to small values anymore. Correspondingly,
the Green tensor in Eq.~(\ref{L60-2}) is the one that takes
the presence of the dielectric body fully into account,
whereas in the Green tensor in Eq.~(\ref{L8}) the presence 
of the dielectric body is not considered.
Hence in contrast to Eq.~(\ref{L8}), Eq.~(\ref{L60-2})
includes in the calculation of the Casimir force that acts 
on a dielectric body the body's retroaction on the 
electromagnetic ground-state noise of the residual system.

To make this explicit, one can expand the full Green tensor
$\tensor{G}_{V}(\mathbf{r,r'},i\xi)$
in powers of $\bdelta \varepsilon (\mathbf{r},i\xi)$
by using the iterative solution to Eq.~(\ref{L14a}) 
[together with Eq.~(\ref{L14c})]. Inserting the resulting Born
series for $\tensor{G}_{V}(\mathbf{r,r'},i\xi)$
in Eq.~(\ref{L60-1}), one obtains the corresponding expansion of the
Casimir force $\mathbf{F}$ in powers of 
$\bdelta\varepsilon(\mathbf{r},i\xi)$. In particular, 
truncating this expansion at the term linear
in $\bdelta\varepsilon(\mathbf{r},i\xi)$
$[\bdelta\varepsilon(\mathbf{r},i\xi)
\mapsto\delta\varepsilon(\mathbf{r},i\xi)$], i.e.,
replacing $\tensor{G}_{V}(\mathbf{r,r'},i\xi)$
with its zeroth-order approximation
$\tensor{G}(\mathbf{r,r'},i\xi)$, we simply obtain
\begin{align}
\label{L61-1}
&
\mathbf{F}
=-\frac{\hbar}{2\pi c^2}
\int_{0}^{\infty} \D\xi\,
\xi^2
\nonumber\\&\quad \times\,\biggl\{
\int_{V_\mathrm{M}} \D^3r\,
\chi
_\mathrm{M}
(\mathbf{r},i\xi)\Nabla
\mathrm{Tr\,}[
\tensor{G}(\mathbf{r,r'},i\xi)]
_{\mathbf{r'}\to\mathbf{r}}
\nonumber\\&\qquad\quad
-2\int_{\partial V_\mathrm{M}}
\D\mathbf{a}
\cdot
\chi
_\mathrm{M}
(\mathbf{r},i\xi)
[\tensor{G}(\mathbf{r,r'},i\xi)]
_{\mathbf{r'}\to\mathbf{r}}
\biggr\}.
\end{align}
In this case, the prescription for taking the coincidence
limit of the Green tensor simply consists in the replacement
$[\tensor{G}(\mathbf{r,r'},i\xi)]_{\mathbf{r}'\to\mathbf{r}}
\mapsto \tensor{G}^{\mathrm{(S)}}(\mathbf{r,r},i\xi)$,
where $\tensor{G}^{\mathrm{(S)}}(\mathbf{r,r'},i\xi)$
is the scattering part of the Green tensor
$\tensor{G}(\mathbf{r,r'},i\xi)$ in the absence of the 
dielectric matter in $V$ (cf. Figs.~\ref{Fig1} and \ref{Fig2}).
Note that if the surface integral can be dropped (i.e., if 
the case sketched in Fig.~\ref{Fig1} is considered), then 
Eq.~(\ref{L61-1}) becomes identical with Eq.~(\ref{L8}).
It should be mentioned that inclusion in Eq.~(\ref{L61-1}) of 
the higher-order terms of the Born
series of $\tensor{G}_{V}(\mathbf{r,r'},i\xi)$
generates an increasing number of many-body corrections.

It may be also informative to examine the effect of the change
$\bdelta\varepsilon(\mathbf{r},\omega)$ not only on the level 
of the expectation values in Eqs.~(\ref{L51a}) and (\ref{L51b}), but
more directly on the level of the corresponding operators.
This is outlined in Appendix \ref{AppB} for weakly dielectric 
material [$\bdelta\varepsilon(\mathbf{r},\omega)
\mapsto \delta\varepsilon(\mathbf{r},\omega)$]. In this context, 
an alternative derivation of Eq.~(\ref{L61-1}) is given.

In view of Eq.~(\ref{L45-1}), Eq.~(\ref{L60-1}) can be also
given in the form of a surface integral, where the
stress tensor $\tensor{T}(\mathbf{r})$ can be found by 
studying [in a similar way as in the derivation of Eq.~(\ref{L60-1})]
the change induced in Eqs.~(\ref{L12}) and (\ref{L13})
by the change $\bdelta\varepsilon(\mathbf{r},i\xi)$.
More easily, we may directly derive it (up to irrelevant
transverse contributions) from Eq.~(\ref{L60-1}) as
\begin{align}
\label{L72a}
&\tensor{T}(\mathbf{r})
=
\frac{\hbar}{\pi c^2}
\int_{0}^{\infty} \D\xi\,
\xi^2
\chi
_\mathrm{M}
(\mathbf{r},i\xi)
[\tensor{G}_{V}(\mathbf{r,r},i\xi)]_{\mathbf{r'}\to\mathbf{r}}
\nonumber\\&\qquad\quad
+
\frac{\hbar}{2\pi c^2}\,
\Nabla
\int_{0}^{\infty}\D\xi\,
\xi^2
\int_{V_\mathrm{M}} \D^3r'\,
\nonumber\\&\qquad\qquad
\times \
\frac{
\chi
_\mathrm{M}
(\mathbf{r'},i\xi)
}
{4\pi\,|\mathbf{r-r'}|}\,
\Nabla{'}
\mathrm{Tr\,}
[\tensor{G}_{V}(\mathbf{r'},\mathbf{r''},i\xi)]
_{\mathbf{r''}\to\mathbf{r'}},
\end{align}
which makes obvious the fact that the stress tensor depends
on the permittivity in a spatially non-local way in general.
With respect to the volume integral in Eq.~(\ref{L72a}),
$\mathbf{r}$ should be thought of as being infinitesimally
outside $V_\mathrm{M}$ if necessary.
The formulation of the force in terms of the stress tensor
is particularly advantageous in the case of homogeneous
material. {F}rom Eq.~(\ref{L60-1}) it is easily seen that
for a homogeneous body that is an inner part of a larger
body (cf. Fig.~\ref{Fig2}) the somewhat cumbersome stress 
tensor (\ref{L72a}) can be replaced with
\begin{multline}
\label{L74b}
\tensor{T}
(\mathbf{r})
= - \frac{\hbar}{\pi c^2}
\int_{0}^{\infty} \D\xi\,
\xi^2
\chi
_\mathrm{M}
(i\xi)
\bigl\{
{\textstyle\frac{1}{2}}\tensor{1}
\mathrm{Tr\,}
[\tensor{G}_{V}(\mathbf{r,r'},i\xi)]
\\
- \tensor{G}_{V}(\mathbf{r,r'},i\xi)
\bigr\}_{\mathbf{r'}\to\mathbf{r}},
\end{multline}
and for an isolated homogeneous body (cf. Fig.~\ref{Fig1})
it follows from Eq.~(\ref{L60-2}) that it can be replaced with
\begin{equation}
\label{L74a}
\tensor{T} (\mathbf{r})
=
-\frac{\hbar}{2\pi c^2}
\int_{0}^{\infty}\D\xi\,
\xi^2
\chi
_\mathrm{M}
(i\xi)\tensor{1}
\mathrm{Tr\,}[\tensor{G}_{V}
(\mathbf{r},\mathbf{r'},i\xi)]_{\mathbf{r'}\to\mathbf{r}}.
\end{equation}
Furthermore, the assumed homogeneity then
implies that we may let
$[\tensor{G}_{V}(\mathbf{r,r'},i\xi)]_{\mathbf{r}'\to\mathbf{r}}
\mapsto \tensor{G}_{V}^{\mathrm{(S)}}(\mathbf{r,r},i\xi)$
in Eqs.~(\ref{L74b}) and (\ref{L74a}). Needless to say that
replacing the full Green tensor $\tensor{G}_{V}
(\mathbf{r,r'},i\xi)$ as appearing in Eqs.~(\ref{L72a})--(\ref{L74a})
with the zeroth-order approximation $\tensor{G}(\mathbf{r,r'},i\xi)$ 
yields again the Casimir force in the
case of weakly dielectric material.

%%%%%%%%%%%%%%%%%%%%%%%%%%%%%%%%%%%%%%%%%%%%%%%%%%%%%%%%%%
\section{Force on micro-objects and atoms}
\label{sec5}
Since nothing has been said about the spatial extension of the
bodies under consideration, the applicability of
Eqs.~(\ref{L60-1}) and (\ref{L60-2}) ranges from dielectric
macro-objects to micro-objects, even including single atoms. 
Let us consider a dielectric body that may be thought of
as consisting of distinguishable
(electrically neutral but polarizable)
micro-constituents frequently called atoms or molecules within the
framework of molecular optics. We may then assume the validity of the
Clausius--Mosotti relation \cite{Jackson3rd,BornWolf},
\begin{align}
\label{L90}
\chi_\mathrm{M}(\mathbf{r},\omega)
&
=\varepsilon_{0}^{-1}\eta(\mathbf{r})
\alpha(\omega)
[1-\eta(\mathbf{r})\alpha(\omega)/(3\varepsilon_{0})]^{-1}
\nonumber\\
&= \varepsilon_{0}^{-1} \eta(\mathbf{r})\alpha(\omega)\,
[1+\chi_\mathrm{M}(\mathbf{r},\omega)/3],
\end{align}
where $\alpha(\omega)$ is the polarizability of
a single micro-constituent and $\eta(\mathbf{r})$ the
number density of the micro-constituents (referred to as 
atoms in the following). It is worth noting that 
there is no need here---in contrast to Eq.~(\ref{L6})---to 
regard $\alpha(\omega)$ as being calculated in the lowest 
(non-vanishing) order of perturbation theory 
according to Eq.~(\ref{L5a}).
It can be shown (Appendix \ref{AppD}) that
Eq.~(\ref{L90}) is consistent with the requirement 
that both $\alpha(\omega)$ and $\chi_\mathrm{M}(\mathbf{r},\omega)$ 
be Fourier transforms of response functions iff 
\begin{equation}
\label{L100}
\eta(\mathbf{r})\alpha(0)/(3\varepsilon_{0}) < 1.
\end{equation}

\subsection{Isolated micro-object}
\label{sec5a}
Let $V_\mathrm{M}$ be the small volume of an isolated dielectric
micro-object (cf. Fig.~\ref{Fig1}) which a dielectric susceptibility 
$\chi_\mathrm{M}(\omega)$ of Clausius--Mosotti-type can be ascribed
to. Combining Eq.~(\ref{L60-2}) with Eq.~(\ref{L90}) and assuming that,
due to the smallness of $V_\mathrm{M}$, the scattering part of the
Green tensor
can be taken out of the space integral at the (appropriately chosen)
position $\mathbf{r}$ of the micro-object,
we derive the force acting on the micro-object to be
\begin{multline}
\label{L90c}
\mathbf{F}
=  - V_\mathrm{M}\eta\,\frac{\hbar\mu_{0}}{2\pi}
\int_{0}^{\infty} \D\xi\, \xi^2\,
\alpha(i\xi)
\left[
1+{\textstyle\frac{1}{3}}\chi_\mathrm{M}(i\xi)
\right]
\\ \times\,
\Nabla
\mathrm{Tr}\,
\tensor{G}_{V}^{\mathrm{(S)}}(\mathbf{r,r},i\xi)
.
\end{multline}
Recall that in the case under study the replacement
$[\tensor{G}_{V}(\mathbf{r,r'},i\xi)]_{\mathbf{r}'\to\mathbf{r}}
\mapsto \tensor{G}_{V}^{\mathrm{(S)}}(\mathbf{r,r},i\xi)$
can be made. Equation (\ref{L90c}), which generalizes Eq.~(\ref{L8b}),
differs in two respects from Eq.~(\ref{L8b}). Firstly, its validity is
no longer restricted to weakly dielectric matter. Secondly, 
it takes into account the dependence of the force on the 
shape of the micro-object.

In contrast to Eq.~(\ref{L8b}), the force as given by Eq.~(\ref{L90c})
includes all-order multi-atom van der Waals interactions 
of the micro-object, as may be seen by expanding the Green tensor 
$\tensor{G}_{V}(\mathbf{r}, \mathbf{r}',i\xi)$ in 
powers of $\chi_\mathrm{M}(i\xi)$ (cf. Ref.~\cite{BuhmannSY102005}).
If they are disregarded, Eq.~(\ref{L90c}) reduces to
(Appendix \ref{AppE})
\begin{equation}
\label{L80}
\mathbf{F}
=
- V_\mathrm{M}\eta\,\frac{\hbar\mu_{0}}{2\pi}
\int_{0}^{\infty} \D\xi\, \xi^2\,
\alpha(i\xi)
\Nabla
\mathrm{Tr}\,
\tensor{G}^{\mathrm{(S)}}(\mathbf{r,r},i\xi)
,
\end{equation}
which, as expected, is nothing but Eq.~(\ref{L8b})---%
only the term linear in $\alpha(i\xi)$ contributes to the force.
The force in this limit is simply the sum of the 
forces acting on the atoms due to the presence of
the external bodies (region $V_\mathrm{B}$ in Fig.~\ref{Fig1}).
Hence, $\mathbf{F}^\mathrm{(at)}$ $\!=$ $\!(V_\mathrm{M}\eta)^{-1}
\mathbf{F}$ is the force acting on a single ground-state atom, 
that is to say, we are left exactly with the formula for 
the CP force as given by Eq.~(\ref{L6}),
with the exception that now the atomic polarizability is the
exact one rather than the perturbative expression given
in Eq.~(\ref{L5b}).

\subsection{Micro-object that is an inner part of a larger body}
\label{sec5b}
Let now $V_\mathrm{M}$ be the small volume of a dielectric
micro-object that belongs to a larger body of volume $V$
of the same atoms (cf. Fig.~\ref{Fig2}).
Under assumptions analogous to those 
leading from Eq.~(\ref{L60-2}) to Eq.~(\ref{L90c}), from Eq.~(\ref{L60-1})
[together with Eq.~(\ref{L90})] we obtain the following formula for
the (shape-dependent) force acting on the micro-object:
\begin{multline}
\label{L90b}
\mathbf{F}
=  - V_\mathrm{M}\eta\,\frac{\hbar\mu_{0}}{\pi}
\int_{0}^{\infty} \D\xi\, \xi^2\,
\alpha(i\xi)
\left[
1+ {\textstyle\frac{1}{3}}\chi_\mathrm{M}(i\xi)
\right]
\\ \times\,
\Nabla\cdot\left[
{\textstyle\frac{1}{2}}\tensor{1}
\mathrm{Tr}\,
\tensor{G}_{V}^{\mathrm{(S)}}(\mathbf{r,r},i\xi)
- \tensor{G}_{V}^\mathrm{(S)}(\mathbf{r,r},i\xi)
\right].
\end{multline}
Equation (\ref{L90b}) differs from Eq.~(\ref{L90c})
in the second term in the square brackets in the second line.
This difference can be regarded as reflecting the fact 
that---in contrast to Eq.~(\ref{L90c})---the force acting 
on the micro-object is screened by the residual part of the body.

When the multi-atom van der Waals interactions of the body
(of volume $V$) are disregarded, then, in a way quite similar to 
that outlined in the derivation of Eq.~(\ref{L80}) in App.~\ref{AppE},
Eq.~(\ref{L90b}) can be shown to reduce to the term linear 
in the atomic polarizability,    
\begin{multline}
\label{L90d}
\mathbf{F}
=  - V_\mathrm{M}\eta\,\frac{\hbar\mu_{0}}{\pi}
\int_{0}^{\infty} \D\xi\, \xi^2\,
\alpha(i\xi)
\\ \times\,
\Nabla\cdot\left[
{\textstyle\frac{1}{2}}\tensor{1}
\mathrm{Tr}\,
\tensor{G}^{\mathrm{(S)}}(\mathbf{r,r},i\xi)
- \tensor{G}^\mathrm{(S)}(\mathbf{r,r},i\xi)
\right]
.
\end{multline}
Recall that in this approximation
$\tensor{G}_{V}^{\mathrm{(S)}}(\mathbf{r,r},i\xi)$
can be replaced with $\tensor{G}^{\mathrm{(S)}}(\mathbf{r,r},i\xi)$.
{F}rom Eq.~(\ref{L90d}) it then follows that
$\mathbf{F}^\mathrm{(at)}$ $\!=$ $\!(V_\mathrm{M}\eta)^{-1}
\mathbf{F}$ can be regarded as the
screened CP force acting on an atom of a weakly dielectric medium.

To make contact with earlier results, let us apply
Eq.~(\ref{L90d}) to the atoms of a weakly dielectric medium
(corresponding to the region $V$ in Fig.~\ref{Fig2})
in front of a laterally infinitely extended
magnetodielectric planar wall
(corresponding to the region $V_\mathrm{B}$ in Fig.~\ref{Fig2}),
which is assumed to extend from some negative
$z$\,value up to $z$ $\!=$ $\!0$. Using the
explicit form of the Green tensor for planar multi-layer
structures (see, e.g., \cite{TomasMS001995,ChewBook}),
we may write its scattering part for coincident spatial 
arguments in the (empty) space region \mbox{$z$ $\!>$ $\!0$} as
\begin{multline}
\label{A1}
\tensor{G}^{\mathrm{(S)}}(\mathbf{r,r},\omega)=
\frac{i}{8\pi^2k^2}\int \D^2q\,\frac{e^{2i\beta z}}{\beta}
\,\bigl\{
r_{-}^{p}
\bigl [q^2 \uv{z}\uv{z}
\\
-\beta^2 \uv{q}\uv{q}\bigr]
%\\
+r_{-}^{s}k^2 \uv{s}\uv{s}
\bigr\},
\end{multline}
with $k$ $\!=$ $\!k(\omega)$ $\!=$ $\!\omega/c$, $q$ $\!=$
$\!|\mathbf{q}|$,
$\beta$ $\!=$ $\!\beta(\omega,q)$ $\!=$ $\!(k^2-q^2)^{1/2}$ 
and orthogonal unit vectors $\uv{q}$ $\!=$ $\!\mathbf{q}/q$, 
$\uv{z}$ $\!=$ $\!\Nabla z$, and $\uv{s}$ $\!=$ $\!\uv{q}$ 
$\!\times$ $\!\uv{z}$. The effect of the (multi-layered) wall 
is described in terms of the generalized reflection coefficients
$r_{-}^{\sigma}$ $\!=$ $\!r_{-}^{\sigma}(\omega,q)$
($\sigma$ $\!=$ $\!s,p$), which in the simplest case of an 
internally homogeneous, semi-infinite wall reduce to the 
usual Fresnel amplitudes. From Eq.~(\ref{A1}) it then follows that
\begin{equation}
\label{A2}
\mathrm{Tr\,}
\tensor{G}^{\mathrm{(S)}}(\mathbf{r,r},\omega)=
\frac{i}{4\pi}\int_{0}^{\infty} \D q\,q \,\frac{e^{2i\beta z}}{\beta}
\bigl[ (r_{-}^{s}-r_{-}^{p})+\frac{2q^2}{k^2}\,r_{-}^{p}
\bigr].
\end{equation}
Substitution of Eq.~(\ref{A2}) into Eq.~(\ref{L80}) leads to
the well-known expression
\cite{McLachlanAD001963,AgarwalGS001975,WylieJM001984,
HenkelC002002,BuhmannSY032004}
for the CP force acting on a single ground-state
atom in front of a planar wall.
The screened force acting on a weakly dielectric medium atom 
is obtained by substituting Eqs.~(\ref{A1}) and (\ref{A2})
into Eq.~(\ref{L90d}). The result is ($\beta$ $\!=$ $\!i\kappa$)
\begin{align}
\label{A7}
&
\mathbf{F}^{\mathrm{(at)}}(z)
=
(V_\mathrm{M}\eta)^{-1}\mathbf{F}(z)
\nonumber\\&\hspace{2ex}
=
\uv{z}
\frac{\hbar\mu_{0}}{
4
\pi^2}
\!
\int_{0}^{\infty}
\!
\D \xi\,\xi^2\alpha(i\xi)
\!
\int_{0}^{\infty}
\!
\D q\,
q
e^{-2\kappa z}
 (r_{-}^{s}-r_{-}^{p}).
\end{align}
It fully agrees with the result found by calculating
the Casimir stress (\ref{L12}) [together with Eq.~(\ref{L13})] 
in a dielectric layer of a planar multi-layer structure and 
performing therein the limit to weakly dielectric matter
\cite{TomasMS092005,RaabeC112005}.

\section{van der Waals interaction between two atoms}
\label{sec6}
Equation (\ref{L60-1}) can also be regarded as a basic equation
for calculating the force between two (ground-state) atoms.
For this purpose, let us consider the small change 
$\delta\mathbf{F}$ of $\mathbf{F}$ in Eq.~(\ref{L60-1}) due to
a small change $\delta \chi_{1}(\mathbf{r},i\xi)$
of the susceptibility $\chi_{\mathrm M}(\mathbf{r},i\xi)$
and a small change $\delta \chi_{2}(\mathbf{r},i\xi)$
of the susceptibility $\chi_\mathrm{B}(\mathbf{r},i\xi)$
(of one of the bodies) in the region $V_\mathrm{B}$
in Fig.~(\ref{Fig2}). In particular let us assume that
$\chi_\mathrm{M}(\mathbf{r},i\xi)$ only changes 
inside the region $V_\mathrm{M}$.
Recalling Eq.~(\ref{L14a}), it is not difficult to calculate
$\delta\mathbf{F}$ up to second order in
$\delta\chi_k(\mathbf{r},i\xi)$ ($k$ $\!=$ $\!1,2$) and pick
out the
term $\delta_{12}\mathbf{F}$ that is bilinear in
$\delta \chi_{1}(\mathbf{r},i\xi)$ and
$\delta \chi_{2}(\mathbf{r},i\xi)$:
\begin{align}
\label{L60-11}
&
\delta_{12}\mathbf{F}
=\frac{\hbar
}{2\pi
c^4
}\int_{0}^{\infty} \D\xi\,
\xi^4
\int_{V
_\mathrm{M}
} \D^3r\,
\delta
\chi_{1}
(\mathbf{r},i\xi)
\nonumber\\&\quad \times\,
\int_{V_\mathrm{B}} \D^3s\,
\delta\chi_{2}
(\mathbf{s},i\xi)
\Nabla
\mathrm{Tr\,}[
\tensor{G}_{V}(\mathbf{r,s},i\xi)\cdot
\tensor{G}_{V}(\mathbf{s,r},i\xi)
],
\end{align}
where the Green tensor
$\tensor{G}_{V}(\mathbf{r}_{1},\mathbf{r}_{2},i\xi)$
refers to the system before the susceptibilities have been changed.
Note that, since we are dealing with the
interaction between two well-separated space regions,
the problem of removing ``self''-force contributions
does not arise here.

Now let us suppose that the small changes
$\delta\chi_{1}(\mathbf{r},i\xi)$ and
$\delta\chi_{2}(\mathbf{r},i\xi)$ result from
the introduction into the system of additional atoms, say
impurity atoms, of type $1$ and type $2$, respectively.
The (body-assisted) force acting on a type-$1$ atom 
at position $\mathbf{r}_{1}$ due to its interaction with a 
type-$2$ atom at position $\mathbf{r}_{2}$
is then evidently obtained, in first order of their
polarizabilities $\alpha_{1}(i\xi)$ and $\alpha_{2}(i\xi)$,
from the ``crossing term'' $\delta_{12}\mathbf{F}$ as
\begin{multline}
\label{L60-12}
\mathbf{F}_{12}^\mathrm{(at)}
=\frac{\hbar
\mu_{0}^{2}
}{2\pi
}\int_{0}^{\infty} \D\xi\,
\xi^4
\alpha_{1}(i\xi)\alpha_{2}(i\xi)
\\\times\,
\Nabla_{1}
\mathrm{Tr\,}[
\tensor{G}_{V}(\mathbf{r}_{1},\mathbf{r}_{2},i\xi)\cdot
\tensor{G}_{V}(\mathbf{r}_{2},\mathbf{r}_{1},i\xi)
]
,
\end{multline}
which is in full agreement with previous calculations of
the van der Waals interaction between two atoms
\cite{MahantyJ001972,MahantyJ001973,MahantyBook}.
Recall that $\tensor{G}_{V}(\mathbf{r}_{1},\mathbf{r}_{2},i\xi)$ 
is the Green tensor for the material system that was present 
before the introduction of the additional atoms.

Disregarding local-field corrections, one may insert 
in Eq.~(\ref{L60-12}) the Green tensor for the unperturbed 
host media. In particular, the force between two atoms embedded 
in a homogeneous (dielectric) background medium is then obtained by
identifying $\tensor{G}_{V}(\mathbf{r}_{1},\mathbf{r}_{2},i\xi)$
with the well-known bulk-medium Green tensor.
Note that in this case the same formula for the force
can be obtained by basing the calculations on the
Minkowski stress tensor
\cite{MarcowitchM112005,TomasMS092005,AbrikosovStatPhys}.
Choosing in Eq.~(\ref{L60-12}) the free-space Green tensor,
we recover the van der Waals interaction between two atoms in
otherwise empty space. It should be pointed out
that $\mathbf{F}_{12}^\mathrm{(at)}$ and 
$\mathbf{F}_{21}^\mathrm{(at)}$ obey the lex tertia
$\mathbf{F}_{12}^\mathrm{(at)}$ $\!=$ 
$\!-\mathbf{F}_{21}^\mathrm{(at)}$ if
the Green tensor is translationally invariant
[$\tensor{G}_{V}(\mathbf{r}_{1}$ $\!
+$ $\!\mathbf{v},\mathbf{r}_{2}$ $\!
+$ $\!\mathbf{v},i\xi)$ $\!=$ $\!\tensor{G}_{V}(\mathbf{r}_{1},
\mathbf{r}_{2},i\xi)$],
as it is the case for the two atoms being in bulk material or
in free space. Since Eq.~(\ref{L60-12}) describes the 
atom--atom force in the presence of
arbitrary macroscopic bodies, it is clear that
the atomic positions $\mathbf{r}_{1}$ and
$\mathbf{r}_{2}$ are not physically equivalent in general.

\section{Summary and Conclusions}
\label{sec7}
Within the framework of macroscopic QED in linearly, locally, and
causally responding media,
we have shown that dispersive forces acting on (ground-state)
macro- and micro-objects---including single atoms---can be
calculated in a unified way on the basis of the Lorentz force 
density that acts on the charge and current
densities attributed to the polarization and magnetization of the
media. Although the examples considered in Secs.~\ref{sec4}--\ref{sec6}
refer to dielectric objects, the basic formulas given in
Sec.~\ref{sec3} can also be used to include in the calculations
magnetic properties of the matter. Inclusion in the theory
of non-locally responding
media would require an extension of the underlying quantization 
scheme, which may be a subject of further studies.

We have derived very general formulas for the force acting on 
a dielectric body or a part of it---formulas which apply to
arbitrary geometries and whose validity is not restricted to
weakly dielectric matter. For locally responding dielectric
matter that may be regarded as consisting of atoms in the 
broadest sense of the word, the permittivity can be assumed
to be of Clausius--Mosotti-type. In this way, all relevant
many-atom van der Waals interactions of the involved matter can
be included in the force to be calculated.

As already mentioned, the applicability of the theory ranges from
macro-objects to micro-objects. Commonly, the force acting on 
a (weakly) dielectric micro-object is calculated in the spirit 
of a simple superposition of CP forces acting on independent atoms.
The present theory enables one to systematically include in 
the calculation both the dependence of the force on the shape 
of the micro-object and, at the same time, the contributions 
to the force due to many-atom interactions of atoms of the 
micro-object, without restriction to weakly dielectric matter.

If the micro-object reduces to a single atom, the well-known formula
for the CP force on a single atom is recovered. It is worth noting that
not only the force acting on an isolated atom can be obtained, but
also the force on a medium atom. For a medium atom, the CP force 
is screened due to the presence of neighboring medium atoms, while there
is of course no such screening in the case of an isolated atom.
Moreover, the basic formulas can also be used to study the
body-assisted van der Waals interaction between atoms.

%%%%%%%%%%%%%%%%%%%%%%%%%%%%%%%%%%%%%%%%%%%%%%%%%%%%%%%%%%%%%%%%%%%
\acknowledgments
We acknowledge fruitful discussions with S.Y. Buhmann. This work was
supported by the Deutsche Forschungsgemeinschaft.
C.R. is grateful for having been granted a Th\"{u}ringer
Landesgraduiertenstipendium.

\begin{appendix}
\section{Derivation of Eqs.~(\ref{L41})-(\ref{L51})}
\label{AppA}
To express $\fo{\rho}(\mathbf{r},\omega)$ and
$\fo{j}(\mathbf{r},\omega)$ as defined by
Eqs.~(\ref{L40a}) and (\ref{L40b}), respectively, in terms of
$\fo{j}_\mathrm{N}(\mathbf{r},\omega)$,
we first insert Eqs.~(\ref{L43}) and (\ref{L44})
in Eqs.~(\ref{L40a}) and (\ref{L40b}). Taking into account
that the Green tensor obeys the differential equation
\begin{equation}
\label{2.1}
\Nabla\times\kappa(\mathbf{r},\omega)\!\Nabla\times
\tensor{G}(\mathbf{r,r'},\omega)
-
\frac{\omega^2}{c^2}\varepsilon(\mathbf{r},\omega)
\!\tensor{G}(\mathbf{r,r'},\omega)
\!=\!\tensor{1}\delta(\mathbf{r\!-\!r'})
\end{equation}
(together with the boundary condition at infinity)
as well as the relation that follows by taking the 
divergence of Eq.~(\ref{2.1}),
\begin{equation}
\label{2.1a}
\frac{\omega^2}{c^2}\,\Nabla\cdot
[\varepsilon(\mathbf{r},\omega)
\tensor{G}(\mathbf{r,r'},\omega)]
=- \Nabla \delta(\mathbf{r\!-\!r'}),
\end{equation}
we easily see by straightforward calculation that
Eqs.~(\ref{L41}) and (\ref{L42}) hold. According to 
Ref.~\cite{KnoellL002001}, the noise current expressed in 
terms of the noise polarization and the noise magnetization,
\begin{equation}
\label{2.4}
\fo{j}_\mathrm{N}(\mathbf{r},\omega) =
-i\omega \fo{P}_\mathrm{N}(\mathbf{r},\omega)
+\Nabla\times\fo{M}_\mathrm{N}(\mathbf{r},\omega),
\end{equation}
can be related to bosonic fields $\hat{\mathbf{f}}_\lambda
(\mathbf{r},\omega)$ and $\hat{\mathbf{f}}_\lambda^\dagger
(\mathbf{r},\omega)$ \mbox{($\lambda$ $\!=$ $\!e,m$)},
\begin{equation}
\label{2.4-3}
\bigl[\hat{f}_{\lambda k}(\mathbf{r},\omega),
\hat{f}_{\lambda'l}^\dagger(\mathbf{r'},\omega')\bigr]
= \delta_{kl}\delta_{\lambda\lambda'}\delta(\mathbf{r}-\mathbf{r}')
\delta(\omega-\omega'),
\end{equation}
by means of the relations
\begin{align}
\label{2.4-1}
&\fo{P}_\mathrm{N}(\mathbf{r},\omega)
=i\left[\hbar\varepsilon_{0}
\Im\varepsilon(\mathbf{r,\omega})/\pi\right]^{1/2}
\hat{\mathbf{f}}_{e}(\mathbf{r},\omega),
\\
\label{2.4-2}
&\fo{M}_\mathrm{N}(\mathbf{r},\omega)
=\left[-\hbar\kappa_{0}
\Im\kappa(\mathbf{r,\omega})/\pi\right]^{1/2}
\hat{\mathbf{f}}_{m}(\mathbf{r},\omega)
.
\end{align}
Note that the $\hat{\mathbf{f}}_\lambda
(\mathbf{r},\omega)$
and $\hat{\mathbf{f}}_\lambda^\dagger(\mathbf{r},\omega)$
play the role of the basic variables of the combined system 
composed of the electromagnetic field and the (linear) medium.
Inserting Eqs.~(\ref{2.4-1}) and (\ref{2.4-2}) in Eq.~(\ref{2.4})
and making use of Eq.~(\ref{2.4-3}) we find
that $\fo{j}_\mathrm{N}(\mathbf{r},\omega)$ and
$\fo{j}_\mathrm{N}{^{\hspace{-1.2ex}\dagger}}\,(\mathbf{r},\omega)$
obey the commutation relation
\begin{align}
\label{C3}
&\bigl[\,\hat{\!\underline{j}}_\mathrm{Nk}(\mathbf{r},\omega),
\,\hat{\!\underline{j}}_\mathrm{Nl}{^{\hspace{-1.8ex}\dagger}}\;
(\mathbf{r'},\omega')\bigr]
=\frac{\hbar}{\mu_{0}\pi}\,\delta(\omega-\omega')
\nonumber\\&\quad
\times \,
\left[
\frac{\omega^2}{c^2}
\sqrt{\Im \varepsilon(\mathbf{r},\omega)}\,
\tensor{1}\delta(\mathbf{r\!-\!r'})
\sqrt{\Im \varepsilon(\mathbf{r}',\omega')}
\right.
\nonumber\\&\quad
\left.
+\,\Nabla\!\times\!
\sqrt{\Im \kappa(\mathbf{r},\omega)}\,
\tensor{1}\delta(\mathbf{r\!-\!r'})
\sqrt{\Im \kappa(\mathbf{r}',\omega')}\,
\!\times\!\Lnabla{'}
\right]
_{\!kl}
\!
,
\end{align}
which immediately implies the expression for
the ground-state expectation value
$\EW{\fo{j}_{\mathrm{N}}(\mathbf{r},\omega)
\fo{j}_{\mathrm{N}}^{\dagger}(\mathbf{r'},\omega')}$
as given in Eq.~(\ref{L51}).

%%%%%%%%%%%%%%%%%%%%%%%%%%%%%%%%%%%%%%%%%%%%%%%%%%%%%%%%%%%%%
\section{Alternative derivation of Eq.~(\ref{L61-1})}
\label{AppB}
Let us consider a small variation
$\delta\varepsilon(\mathbf{r},\omega)$ and the corresponding
(first-order) changes of the operators
$\fo{\rho}(\mathbf{r},\omega)$,
$\fo{j}(\mathbf{r},\omega)$,
$\fo{E}(\mathbf{r},\omega)$, and
$\fo{B}(\mathbf{r},\omega)$,
and assume that $\delta\varepsilon(\mathbf{r},\omega)$ is 
different from zero only in the volume $V$ [cf. Fig.~\ref{Fig2}].
{F}rom Eqs.~(\ref{L40a})-(\ref{L44}) it follows that
\begin{multline}
\label{L61}
\delta\fo{\rho}(\mathbf{r},\omega) =
-\varepsilon_{0}
\Nabla\cdot\{\delta\varepsilon(\mathbf{r},\omega)\fo{E}(\mathbf{r},\omega)
\\
+
[\varepsilon(\mathbf{r},\omega)-1]\delta\fo{E}(\mathbf{r},\omega)
\}
+ (i\omega)^{-1} \Nabla\cdot\delta\fo{j}_\mathrm{N}(\mathbf{r},\omega),
\end{multline}
\begin{align}
\label{L62}
\delta\fo{j}(\mathbf{r},\omega)
=& -i\omega \varepsilon_{0}
\{
\delta\varepsilon(\mathbf{r},\omega)
\fo{E}(\mathbf{r},\omega)
\nonumber\\
& +[\varepsilon(\mathbf{r},\omega)-1]
\delta\fo{E}(\mathbf{r},\omega)
\}
+\delta\fo{j}_\mathrm{N}(\mathbf{r},\omega),
\end{align}
\begin{multline}
\label{L63}
\delta\fo{E}(\mathbf{r},\omega)
=i\mu_{0}\omega\int \D^3s\,
\bigl[
\tensor{G}(\mathbf{r,s},\omega)
\cdot
\delta\fo{j}_{\mathrm{N}}(\mathbf{s},\omega)
\\
+\delta\tensor{G}(\mathbf{r,s},\omega)
\cdot
\fo{j}_{\mathrm{N}}(\mathbf{s},\omega)
\bigr]
,
\end{multline}
\begin{multline}
\label{L64}
\delta\fo{B}(\mathbf{r},\omega)
=\mu_{0}\Nabla\times \int \D^3s\,
\bigl[
\tensor{G}(\mathbf{r,s},\omega)
\cdot
\delta\fo{j}_{\mathrm{N}}(\mathbf{s},\omega)
\\
+\delta\tensor{G}(\mathbf{r,s},\omega)
\cdot
\fo{j}_{\mathrm{N}}(\mathbf{s},\omega)
\bigr].
\end{multline}
Since the state space attributed to the
dynamical variables $\hat{\mathbf{f}}_\lambda
(\mathbf{r},\omega)$ and
$\hat{\mathbf{f}}^\dagger_\lambda(\mathbf{r},\omega)$,
in terms of which the electromagnetic quantities
are thought of as being expressed,
can be regarded as being independent of the chosen
permittivity (and/or permeability)
\cite{KnoellL002001}, we may apply
the rule $\EW{\delta\cdots}$ $\!=$ $\!\delta\EW{\cdots}$
when calculating expectation values.
Thus, combining Eqs.~(\ref{L40a})-(\ref{L44}) with
Eqs.~(\ref{L61})-(\ref{L64}), we derive
\begin{align}
\label{L65}
&\delta\EW{\rho(\mathbf{r},\omega)
\fo{E}^{\dagger}(\mathbf{r'},\omega')}= -\varepsilon_{0}
\Nabla\!\cdot\![\delta\varepsilon(\mathbf{r},\omega)
\EW{\fo{E}(\mathbf{r},\omega)
\fo{E}^{\dagger}(\mathbf{r'},\omega')}]
\nonumber\\&\  +
\frac{1}{i\omega}\,
\Nabla\cdot
[
\EW{
\delta\fo{j}_\mathrm{N}(\mathbf{r},\omega)
\fo{E}^{\dagger}(\mathbf{r'},\omega')
}+
\EW{
\fo{j}_\mathrm{N}(\mathbf{r},\omega)
\delta\fo{E}^{\dagger}(\mathbf{r'},\omega')}]
\nonumber\\&\
+\ldots
\end{align}
and
\begin{align}
\label{L66}
&\delta\EW{\fo{j}(\mathbf{r},\omega)
\fo{B}^{\dagger}(\mathbf{r'},\omega')
}=
-i\omega\varepsilon_{0}
%[
\delta\varepsilon(\mathbf{r},\omega)
\EW{\fo{E}(\mathbf{r},\omega)
\fo{B}^{\dagger}(\mathbf{r'},\omega')}
%]
\nonumber\\&\
+
[\EW{
\delta\fo{j}_\mathrm{N}(\mathbf{r},\omega)
\fo{B}^{\dagger}(\mathbf{r'},\omega')
}+
\EW{
\fo{j}_\mathrm{N}(\mathbf{r},\omega)
\delta\fo{B}^{\dagger}(\mathbf{r'},\omega')}]
\nonumber\\&\
+\ldots
,
\end{align}
where terms that vanish when $\mathbf{r},\mathbf{r}'\in V$
have not been quoted [$\varepsilon(\mathbf{r},\omega)$ 
$\!=$ $\!1$ in $V$]. Now from Eq.~(\ref{L45}) together with 
Eqs.~(\ref{L65}) and (\ref{L66}) and Eqs.~(\ref{L43}), (\ref{L44}), 
(\ref{L51}), (\ref{L63}), and (\ref{L64}) we may easily calculate 
$\delta\mathbf{F}$. On recalling standard properties
of the Green tensor, we derive
\begin{equation}
\label{B7-0}
\delta\mathbf{F} = \delta^{(1)}\mathbf{F} + \delta^{(2)}\mathbf{F},
\end{equation}
where
\begin{align}
\label{L67}
&
\delta^{(1)} \mathbf{F}
=\frac{\hbar}{2\pi}\int_{0}^{\infty}\!\!\! \D\omega\, 
\frac{\omega^2}{c^2}
\nonumber\\&\quad \times\,\biggl\{
\int_{V_\mathrm{M}} \!\D^3r\,
\delta\varepsilon(\mathbf{r},\omega)\Nabla
\mathrm{Tr\,}
[
\Im
\tensor{G}(\mathbf{r,'},\omega)
]_{\mathbf{r'}\to\mathbf{r}}
\nonumber\\&\qquad\quad
-2
\int_{\partial V_\mathrm{M}}
\D\mathbf{a}
\cdot
\delta\varepsilon(\mathbf{r},\omega)
\Im
[\tensor{G}(\mathbf{r,r'},\omega)]_{\mathbf{r'}\to\mathbf{r}}
\biggr\}
\end{align}
arises from the first terms on the right-hand sides of
Eqs.~(\ref{L65}) and (\ref{L66}), and 
\begin{align}
\label{L69}
&\delta^{(2)} \mathbf{F}
=\frac{\hbar}{2\pi}\int_{0}^{\infty}\D\omega\,\frac{\omega^2}{c^2}
\nonumber\\&\quad \times\,\biggl\{
\int_{V_\mathrm{M}}\!\! \D^3r\,
[
\Im\delta\varepsilon(\mathbf{r},\omega)]\Nabla
\mathrm{Tr\,}
[\tensor{G}^{\ast}(\mathbf{r,r'},\omega)]_{\mathbf{r'}\to\mathbf{r}}
\nonumber\\&\qquad\quad
-2
\int_{\partial V_\mathrm{M}}
\D\mathbf{a}
\cdot
[\Im\delta\varepsilon(\mathbf{r},\omega)]
[\tensor{G}^{\ast}(\mathbf{r,r'},\omega)]
_{\mathbf{r'}\to\mathbf{r}}
\biggr\}
%.
\end{align}
arises from the second terms on the right-hand sides of
Eqs.~(\ref{L65}) and (\ref{L66}). Note that in the derivation 
of Eq.~(\ref{L69}), the relation
\begin{multline}
\label{L68}
\EW{\delta
\fo{j}_{\mathrm{N}}(\mathbf{r},\omega)
\fo{j}_{\mathrm{N}}^{\dagger}(\mathbf{r'},\omega')
}=
\EW{
\fo{j}_{\mathrm{N}}(\mathbf{r},\omega)
\delta\fo{j}_{\mathrm{N}}^{\dagger}
(\mathbf{r'},\omega')
}
\\
=\frac{\hbar}{2\mu_{0}\pi}\delta(\omega-\omega')
\frac{\omega^2}{c^2}\,
\Im\,\delta\varepsilon(\mathbf{r},\omega)
\tensor{1}\delta(\mathbf{r-r'})
\end{multline}
has been used, which follows from Eq.~(\ref{L51}).
Combining Eqs.~(\ref{B7-0}), (\ref{L67}), and (\ref{L69}) yields
\begin{align}
\label{L70}
&\delta \mathbf{F}
=\frac{\hbar}{2\pi}\int_{0}^{\infty} \D\omega\, \frac{\omega^2}{c^2}\,
\nonumber\\&\quad \times\,\Im\biggl\{
\int_{V_\mathrm{M}} \D^3r\,
\delta\varepsilon(\mathbf{r},\omega)\Nabla
\mathrm{Tr\,}[
\tensor{G}(\mathbf{r,r'},\omega)
]_{\mathbf{r'}\to\mathbf{r}}
\nonumber\\&\qquad\quad
-2
\int_{\partial V_\mathrm{M}}
\D\mathbf{a}
\cdot
\delta\varepsilon(\mathbf{r},\omega)
[\tensor{G}(\mathbf{r,r'},\omega)]_{\mathbf{r'}\to\mathbf{r}}
\biggr\}
.
\end{align}
Changing in Eq.~(\ref{L70}) the real-frequency integral to
an imaginary-frequency integral in the usual way, we just arrive
at Eq.~(\ref{L61-1}) [$\delta\varepsilon(\mathbf{r},i\xi)
\mapsto\chi_\mathrm{M}(\mathbf{r},i\xi)$,  
$\delta\mathbf{F}\mapsto\mathbf{F}$]. Note that only 
the real parts of Eqs.~(\ref{L67}) and (\ref{L69})
contribute to Eq.~(\ref{L70}), while the imaginary parts drop out.

\section{Clausius--Mosotti susceptibility and causality}
\label{AppD}
Since $\alpha(\omega)$ is
the Fourier transform of a
response function, it
is holomorphic and without zeros in the upper 
complex half-plane (see, e.g., Ref.~\cite{LanLif} for a summary of
response function properties).
Consequently, $\chi_\mathrm{M}(\mathbf{r},\omega)$
as given by 
(the first line of)
Eq.~(\ref{L90})
is there also holomorphic and without zeros, except for
possible poles at
$\omega$-values satisfying the equation
$\eta(\mathbf{r})\alpha(\omega)/(3\varepsilon_{0})\!=\!1$.
However, since
$\alpha(\omega)$ is 
the Fourier transform of
a response function,
it is real
only on the imaginary frequency axis,
where it, beginning with
a positive value at
$\omega$ $\!=$ $\!0$, 
monotonically decreases with increasing imaginary frequency.
Hence, if
$\eta(\mathbf{r})\alpha(0)/(3\varepsilon_{0})$ 
%$\!>$ 
$\!\geq$
$\!1$
is valid, a pole is observed and
$\chi_\mathrm{M}(\mathbf{r},\omega)$
would fail to be a response function.
{F}rom the requirement that both $\alpha(\omega)$ and
$\chi_\mathrm{M}(\mathbf{r},\omega)$
be Fourier transforms of response functions, it thus
follows that the condition $\eta(\mathbf{r})\alpha(0)
/(3\varepsilon_{0})$ $\!<$ $\!1$ must be imposed on
Eq.~(\ref{L90}).

\section{Derivation of Eq.~(\ref{L80})}
\label{AppE}
In order to derive Eq.~(\ref{L80}), we return to Eq.~(\ref{L60-2}) 
together with Eq.~(\ref{L90}) and recall that, according to 
Eqs.~(\ref{L14c}) and (\ref{L14a}), the Green tensor
$\tensor{G}_{V}(\mathbf{r,r'},\omega)$ obeys the equation
%
% Integration "uber ganz V, wenn stattdessen
% Situation in Fig2 betrachtet wird:
\begin{multline}
\label{E6}
\tensor{G}_{V}
(\mathbf{r,r'},\omega)=\tensor{G}
(\mathbf{r,r'},\omega)
\\
+\frac{\omega^2}{c^2}
\int
_{V_\mathrm{M}}
\D^3s\,
\tensor{G}
(\mathbf{r,s},\omega)
\cdot
\chi_\mathrm{M}(\mathbf{s},\omega)
\tensor{G}_{V}(\mathbf{s,r'},\omega).
\end{multline}
According to the principal volume method
\cite{vanBladelJ111961}, the decomposition
\begin{equation}
\label{E10}
\tensor{G}(\mathbf{r},\mathbf{s},\omega)
=\mathcal{P}\tensor{G}(\mathbf{r},\mathbf{s},\omega)
-\frac{c^2}{\omega^2}\,\tensor{L}\delta(\mathbf{r-s})
\end{equation}
is applicable to the Green tensor
$\tensor{G}(\mathbf{r},\mathbf{s},\omega)$ in Eq.~(\ref{E6}).
The procedure to be followed is to choose first
an exclusion volume whose shape determines the tensor
$\tensor{L}$, and to subsequently treat integrals
over $\mathcal{P}\tensor{G}(\mathbf{r},\mathbf{r'},\omega)$
as (shape-dependent) principal value integrals over an 
exclusion volume with the specified shape (see, e.g.,
Ref.~\cite{ChewBook} for details).
Using Eq.~(\ref{E10}) in Eq.~(\ref{E6}), and adopting a spherical
exclusion volume for which $\tensor{L}$ $\!=$ $\!\tensor{1}/3$,
%(Derselbe Wert gilt auch fuer ein wuerfelfoermiges exclusion volume)
we have
%
% Integration "uber ganz V, wenn stattdessen
% Situation in Fig2 betrachtet wird:
%
\begin{align}
\label{E11}
&
\tensor{G}_{V}
(\mathbf{r,r'},i\xi)=
[1+\chi_\mathrm{M}(\mathbf{r},i\xi)/3]^{-1}
\bigl[
\tensor{G}
(\mathbf{r,r'},i\xi)\nonumber\\&\; 
-\frac{\xi^2}{c^2}
\int_{V_\mathrm{M}}\! \D^3s\,
\mathcal{P}\tensor{G}(\mathbf{r,s},i\xi)
\cdot
\chi_\mathrm{M}(\mathbf{s},i\xi)
\tensor{G}_{V}(\mathbf{s,r'},i\xi)
\bigr]
.
\end{align}
When inserted in Eq.~(\ref{L60-2}), the second term on the right-hand
side of Eq.~(\ref{E11}) leads to a double integral over
$\mathbf{r}$ and $\mathbf{s}$. Since contributions with
$\mathbf{s}$ $\!=$ $\!\mathbf{r}$ are left out
in the principal value integral, the corresponding part of the 
force is associated with
at least two-atom interactions in
$V_\mathrm{M}$.
Dropping all these multi-atom contributions, we obtain
\begin{align}
\label{E12}
&
\mathbf{F}
=-\frac{\hbar}{2\pi c^2}
\int_{0}^{\infty} \D\xi\,
\xi^2
\nonumber\\&\quad \times\,
\int_{V
_\mathrm{M}
} \D^3r\,
\chi
_\mathrm{M}
(\mathbf{r},i\xi)\Nabla
\mathrm{Tr\,}
\left[
\frac{\tensor{G}(\mathbf{r,r'},i\xi)}{
1+\chi_\mathrm{M}(\mathbf{r},i\xi)/3}
\right]
_{\mathbf{r'}\to\mathbf{r}}.
\end{align}
Applying Eq.~(\ref{E12}) to a micro-object
whose number density of atoms is constant over the small
volume $V_\mathrm{M}$, so that
a position-independent Clausius-Mosotti susceptibility
according to Eq.~(\ref{L90})
can be assigned to it, and replacing
$[\tensor{G}(\mathbf{r,r'},i\xi)]_{\mathbf{r}'\to\mathbf{r}}$
with $\tensor{G}^\mathrm{(S)}(\mathbf{r},\mathbf{r},i\xi)$,
we just arrive at Eq.~(\ref{L80}).
\end{appendix}
\bibliographystyle{apsrev}
\bibliography{CasPolPRABib}
%%%%%%%%%%%%%%%%%%%%%%%%%%%%%%%%%%%%%%%%%%%%%%%
\end{document}